\documentstyle[preprint,tighten,aps]{revtex}
\def\lapprox{\mathrel{\mathop
  {\hbox{\lower0.5ex\hbox{$\sim$}\kern-0.8em\lower-0.7ex\hbox{$<$}}}}}
\def\gapprox{\mathrel{\mathop
  {\hbox{\lower0.5ex\hbox{$\sim$}\kern-0.8em\lower-0.7ex\hbox{$>$}}}}}
\def\mathrm{\mbox}
\def\eg{{\em e.g.}\/}
\def\etal{{\em et al.}\/}

\def\ie{{\em i.e.}\/}

\def\fii{\Phi_{\mbox{\footnotesize{i}}}}

\def\fib{\Phi_{\mbox{\footnotesize{B}}}}

\def\fiissm{\Phi_{\mbox{\footnotesize{i}}}^{SSM}}

\newcommand{\oB}{$^8$B~}

\begin{document}
\preprint{\null\hfill  INFNFE-09-96}
\title{Solar opacity, neutrino signals and
 helioseismology 
%\footnote{Presented at "New Trends in solar
%neutrino physics", Laboratori Nazionali del Gran Sasso (Italy), May 1996, to
%appear in the proceedings of the workshop.}
}

\author{
%G. Fiorentini$^{1,2}$\footnote{fiorentini@vaxfe.fe.infn.it}
 B. Ricci\footnote{ricci@vaxfe.fe.infn.it}}
%\centerline{\em 
\address{
%      $^1$ Dipartimento di Fisica, Universit\`a di Ferrara, via Paradiso 
%      12, I-44100 Ferrara, Italy \\
      Istituto Nazionale di Fisica Nucleare, Sezione di Ferrara, 
      via Paradiso 12, I-44100 Ferrara, Italy 
      }

%\date{}

\maketitle                 % Produces the title.

\begin{abstract}
In connection with the recent suggestion by Tsytovich \etal~\cite{TSY}
that opacity in the solar core could be overestimated, we consider
 the following questions:
           i) What would a 10\% opacity 		
			reduction imply for the solar neutrino 
			puzzle?\\
	     ii) Is there any hope of solving 		
			the solar neutrino puzzle by		
			changing  opacity?\\
		iii) Is a 10\% opacity reduction testable 
			with helioseismological data? \\

\end{abstract}

\vskip1.5truecm

Recently Tsytovich \etal~\cite{TSY} reviewed corrections to the theory of 
photon transport in  a dense hot plasma and claimed that previous calculations
overestimate solar opacity $\kappa$ near solar center by about 10\%.
More generally the accuracy of the calculated 
opacity in the solar core is controversial: the characteristic difference
between the calculations of the Livermore and Los Alamos 
groups is about (2--5)\% \cite{RI91} and essentially on these grounds
  Bahcall and Pinsonneault \cite{BP92} estimate
a ($1\sigma$) uncertainty of about 2.5\%. On the other hand, Turck-Chi\`eze 
\etal \cite{TC88} argued the uncertainty to exceed  5\%, a point criticized
in \cite{BP92}. It is worth recalling that, for the conditions of the 
solar  interior, no experimental check is available and the 
estimates of accuracy  originate essentially from theoretical arguments.

In this short note we consider thus 
 the following questions:

           i) What would 10\% opacity 		
			reduction imply for the solar neutrino 
			puzzle?

	     ii) Is there any hope of solving 		
			the solar neutrino puzzle by		
			changing  opacity?

		iii) Is a 10\% opacity reduction testable 
			with helioseismological data?

In our analysis
we assume a {\underline{uniform}} reduction of 	
opacity, \ie~ $\kappa (\rho,T,...) \rightarrow x\kappa (\rho,T,...) $, where
the factor $x$ is independent of density, temperature and chemical composition 
(other approaches to the effect of solar opacity on neutrino fluxes are reported,
for instance, in refs. \cite{B69,TCL,CDopa}). The assumption of 
uniform opacity variations is clearly  {\em ad hoc}, to restrict the 
possible solar models. Nevertheless, it can be used as a first attempt to 
analyze the effects of the Tsytovich \etal proposal, for the following 
reasons:

\noindent
1) the plasma effect advocated in \cite{TSY} are mainly governed by
the quantity $\delta=c^2 \omega _{pl}^2/ v_T^2 \omega ^2 $, see \cite{TSY} for 
notations. As $\delta \propto n_e/T^3$, it stays approximately constant 
along the solar profile.

\noindent
2) Assuming the plasma effect  to be se same as in the solar center, the  
consequences  are presumably overestimated, an attitude which is useful 
to tell if the solar neutrino puzzle can be solved by changing opacity.

As a first step, we compare the results of solar neutrino experiments
with the predictions of our 
Standard Solar Model  (SSM) and with those of a solar model
 with opacity  reduced by
10\%, see Table \ref{tabmodelli}.  Our Standard Solar Model
is obtained with the most updated version of FRANEC  
where diffusion of Helium and heavy elements 
is included \cite{Ciacio}.

Essentially, with  respect to the SSM,  one has  a 20\% reduction
 in the Chlorine and Kamiokande 
signals, and a 6\% reduction of Gallium signal.
% Note that these variations are consistent 
% with the analytical laws 
% $\fipp=\fippssm \,  x^{-0.11} \, , 
% \fibe=\fibessm \,  x^{1.1}   \, ,
% \ficno=\ficnossm \,  x^{1.7}$ and  
% $\fib=\fibssm \,  x^{2.4} $,
% derived in  \cite{Report}.
%
The comparison with experimental results shows that this reduced opacity model
is still inconsistent with   data
($\chi^2 /d.o.f.\approx 74$).

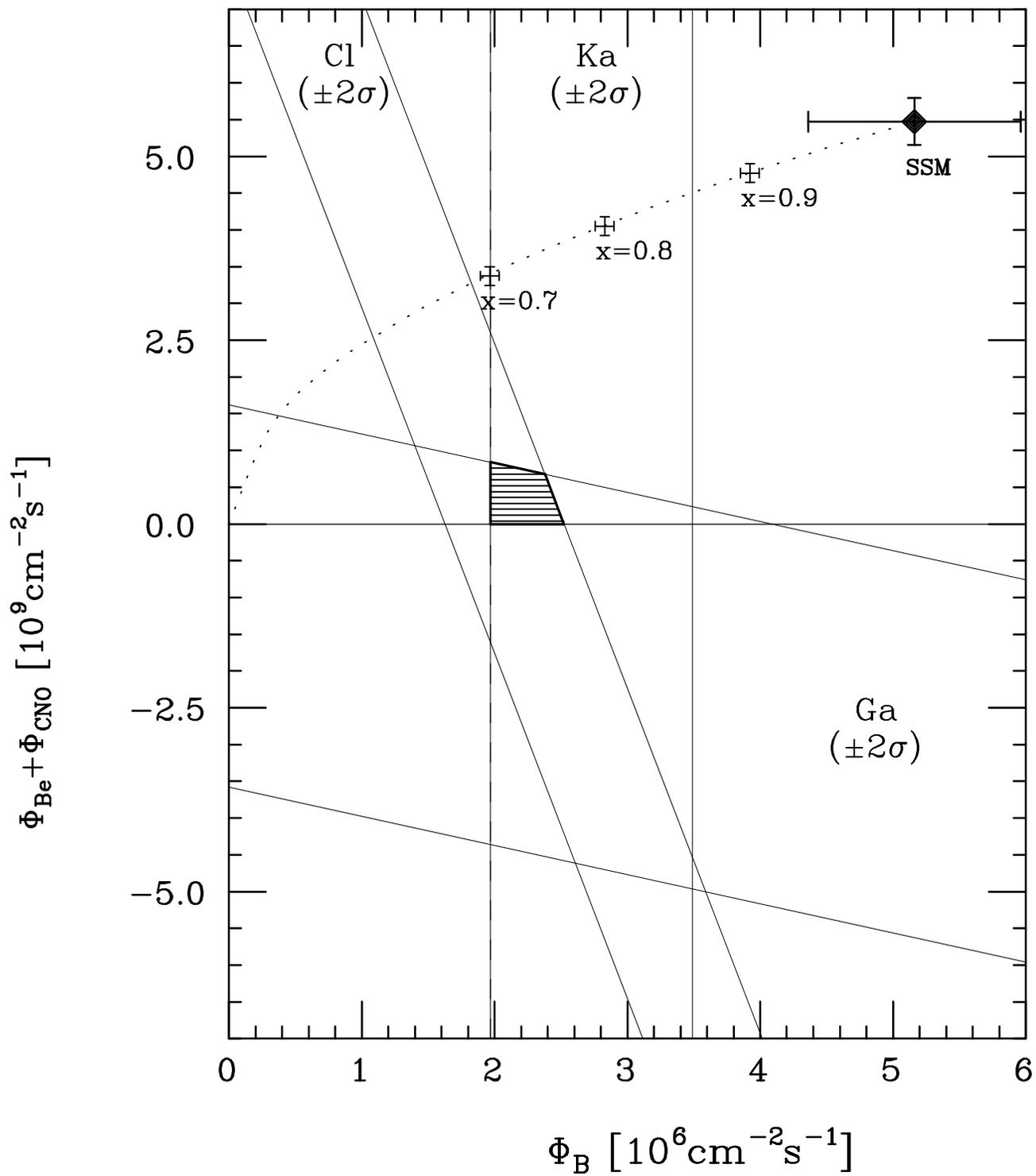
\begin{figure}%[htb]
\caption[fff]{
The  region  within $2\sigma$ from
each experimental result \cite{Ga,Cl,Ka} for standard neutrinos
(dashed area), the prediction of our SSM (diamond) with estimated
$1\sigma$ errors (bars) and the predictions for different opacity reductions
(crosses). The analytical estimate (dotted curve) is also shown.
}
\label{fig2}
\end{figure}

To explore the possibility of solving the solar neutrino puzzle by 
playing with  opacity, we decreased it
down to $x\approx0.6$, see again Tab. \ref{tabmodelli} and  Fig. \ref{fig2}. 
Even a variation of opacity 
 well beyond the theoretical 
uncertainties cannot solve the solar neutrino puzzle:
the best fit is obtained for $x=0.64$ but it corresponds to
$\chi^2 /d.o.f.\approx 10.8$. The reason of the failure is that
 one cannot reproduce simultaneously
the $^7$Be+CNO  and \oB neutrino fluxes implied by experimental results,
in the assumption of standard neutrinos, see again Fig. \ref{fig2}.

To understand what is going on, we remark that
a less opaque Sun is essentially a cooler
Sun \cite{Future,PRD}, with
 practically the same 
profile of density and pressure as the SSM; on the other hand
the temperature profile
has the same shape as that of the SSM, with a changement of scale, by the 
factor:
\begin{equation}
\label{eqt}
T_c/T_c^{SSM}= x^{0.14} \, .
\end{equation}
(These homology relationships are accurate to the 1\%  level or better 
for any quantity characterizing the solar structure, throughout all the radiative
region).

The dependences of neutrino fluxes on central temperature are well known
\cite{PRD,Report,Bahcall1989,BahUlm}, by defining
$\fii=\fiissm \, (T_c/T_c^{SSM})^{\beta_i}$ we found (for our model with 
diffusion):
\begin{equation}
\beta_{p}=-0.7 \quad \beta_{Be}=8.8 \quad \beta_{CNO}=13.6 \quad 
\beta_{B}=19.2 \, .
\end{equation}
By using eq.(\ref{eqt}) one can derive
 the dependence on the opacity parameter $x$. By writing
 $\fii=\fiissm \, x^{\alpha_i}$ one gets:
 \begin{equation}
\alpha_p=-0.1 \quad \alpha_be=1.2 \quad \alpha_{CNO}=1.9  \quad 
\alpha_B=2.7 \, , 
\end{equation}
 which are in excellent agreement with the numerical results.
 % \cite{Report}:
 %\begin{equation}
%\alpha_p=-0.09 \quad \alpha_be=1.2 \quad \alpha_{CNO}=1.9  \quad \alpha_B=2.6 \, . 
%\end{equation}
 
As well known,
 solar models with reduced central temperature cannot
reconcile theory and experiments,
 see \eg~refs. \cite{Future,PRD,Bere96}, and thus it is not a surprise that
 an opacity variation, however large, cannot solve the solar neutrino 
 problem.

In recent years, helioseismology has provided challenging tests of 
the Standard Solar Model.
The values of the Helium mass fraction  and of the mixing length, which were
free parameters for solar model builders before 
the advent of helioseismology, are now strongly constrained by  helioseismological
determinations of the bottom of the convective zone ($R_b$ and $c_b$) and
of the photospheric Helium content ($Y_{photo}$).
By taking into account theoretical uncertainties  (\eg~from equation of 
state and/or different inversion method), helioseismology gives:
\begin{equation}
\label{eqsismo}
R_b/R_\odot=0.710 - 0.716 \quad ; \quad c_b =(0.221 - 0.225)\,{\mbox{Mm/s}} 
\quad ; \quad  Y_{photo}= 0.233-0.268 \,.
\end{equation} 

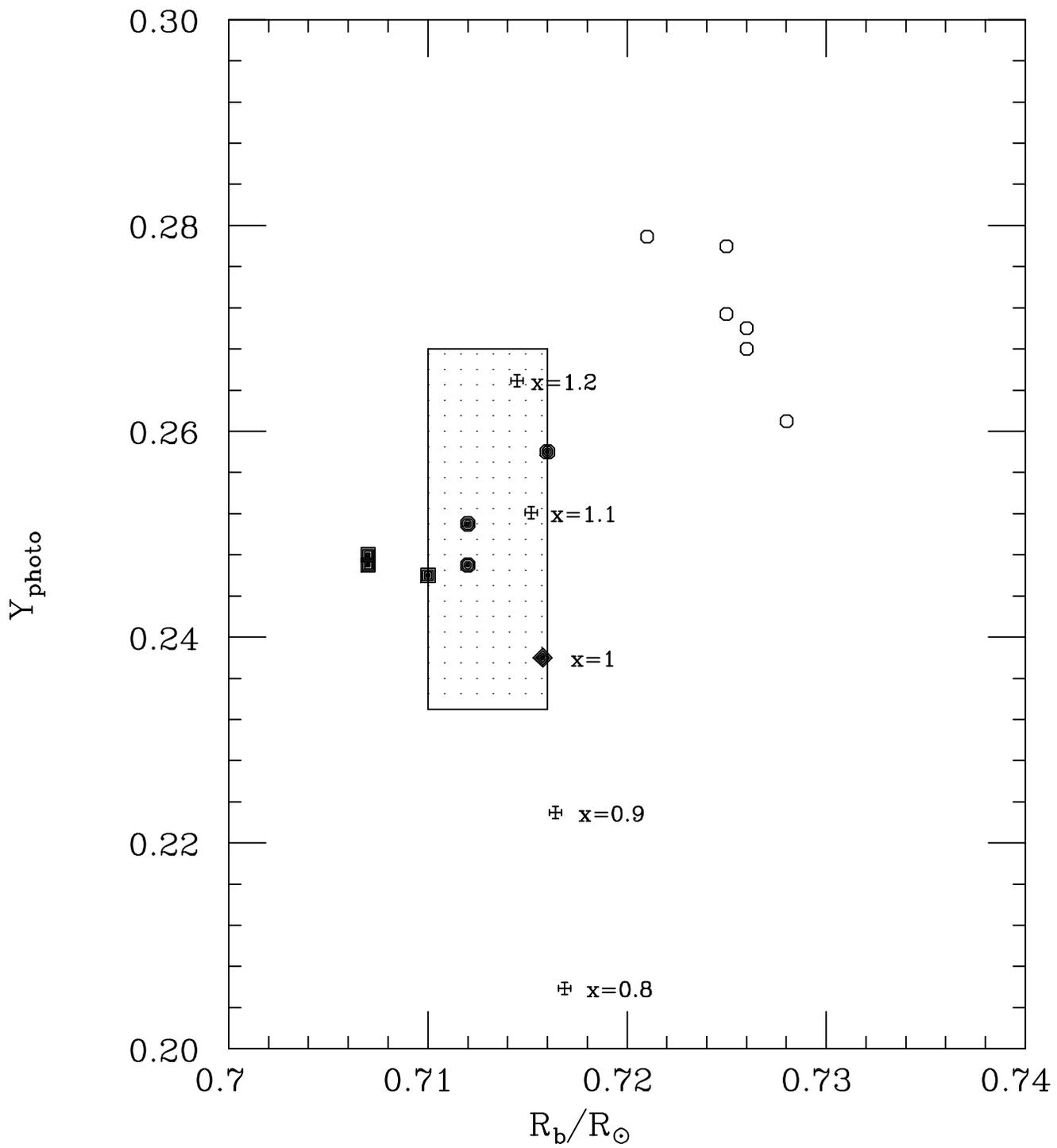
\begin{figure}
\caption[ggg]{The photospheric helium mass fraction $Y_{photo}$ and the 
depth
of the convective zone ($R_b/R_\odot$):\\
a) as constrained from helioseismology (the dotted rectangle)\\
b) as predicted by solar models without diffusion (open circles),
 with helium diffusion  (full squares) and with helium and heavy
 elements diffusion (full circles and diamond). Also shown are the effects
 of varying opacity, for the indicated values of $x$. 
}
\label{figsismo}
\end{figure}

Actually, within the present uncertainties, the information on $R_b$ and
on $c_b$ are not independent, since to the per cent level, $c_b$ is related
to $R_b$ through \cite{sismo}:
\begin{equation}
c_b^2\simeq \frac{2}{3} \frac{G M_\odot}{R_\odot}  \left ( \frac{R_\odot}{R_b} 
-1 \right )
\end{equation}
Concerning $Y_{photo}$ we have reported in eq. (\ref{eqsismo}) the total range
of published helioseismological determinations, see \cite{Report}.
It is important that only standard solar models including diffusion
of helium and heavy elements satisfy these helioseismological constraint,
whereas other models fails, see Tab. \ref{tab2} and Fig. \ref{figsismo}.

Starting from our SSM, we studied the effect of varying the opacity. As 
shown in Fig. \ref{figsismo}, the depth of the convective zone is weakly
sensitive to such variations, whereas the photospheric helium abundance
is sensibly related to opacity. A ten per cent reduction of this latter 
brings $Y_{photo}$ well below the range allowed by helioseismology.

In conclusion, we would like to remark the following points:

i) opacity can be tested to the ten per cent level or better
 by helioseismology (note that we have used only a subset 
 of the helioseismological information, since helioseismology 
  determines the sound
 speed also well below the bottom of the convective zone).
 
 ii) The disagreement between helioseismology and the $x=0.9$ solar
 model is not necessarily a proof against the Tsytovich \etal~proposal,
 as in this case the behaviour of opacity along the solar profile can be 
 crucial.
 
 iii) Last but not least, there is no hope of solving the solar neutrino
 puzzle by playing with opacity.

\acknowledgments
We are extremely grateful to V. Berezinsky and G. Fiorentini for
suggesting the problem as well as for 
their useful 
suggestions, and to J. Bahcall and J. Christensen-Dalsgaard for 
interesting comments.
 We express our gratitude  to S. Degl' Innocenti and F. Ciacio
for the collaborative effort in updating FRANEC.

\begin{table}%[htb]
\caption[ff]{
Experimental results \cite{Ga,Cl,Ka}, predictions of our
SSM and those of solar models with opacity reduced by the multiplicative 
factor $x$ (\ie~$\kappa \rightarrow x \kappa$). The values of 
$\chi^2$ per degree of freedom, calculated
including only experimental errors, are also shown.
}
\begin{tabular}{lccccc}
               &EXP & SSM  & $x=0.9$  & $x=0.8$  & $x=0.7$\\
               \hline
$S_{Cl}$ [SNU] &$2.55\pm0.25$ & 7.4 & 5.9 & 4.5 & 3.4 \\
$S_{Ga}$ [SNU] &$74\pm8$ & 128 & 120 & 113 & 107 \\
$\fib$ [10$^6$/cm/s] &$2.73\pm0.38$ & 5.16 & 3.92 & 2.83 & 1.96  \\
\hline
$\chi^2/d.o.f.$ &        &154 & 74 & 28 & 11\\
\end{tabular}
\label{tabmodelli}
\end{table}

\begin{table}
\caption[gg]{The depth of the convective zone  and
the photospheric helium abundance as determined from
helioseismology and as predicted by recent standard solar models.
}
\begin{tabular}{lcc}
    &$R_b/R_\odot$   &  $Y_{photo}$ \\
 \hline
 Helioseismologhy & 0.710--0.716 & 0.233--0.268 \\
 \hline
 \hline
 \multicolumn{3}{l}{ \bf SSMs with diffusion of He and Z}\\
 FRANEC96 \cite{Ciacio} & 0.716  & 0.238\\
 BP95     \cite{BP95}   & 0.712  & 0.247\\
 P94      \cite{P94}    & 0.712  & 0.251\\
 RVCD96   \cite{RCVD96} & 0.716  & 0.258\\
 \hline
 \hline
  \multicolumn{3}{l}{\bf SSMs with diffusion of He}\\ 
  BP92    \cite{BP92}   & 0.707  & 0.247\\
  P94     \cite{P94}    & 0.710  & 0.246\\
  BCDSTT  \cite{Basu}   & 0.707  & 0.248\\
  \hline
  \hline
   \multicolumn{3}{l}{\bf SSMs without diffusion}\\
    FRANEC96-ND \cite{Ciacio}    & 0.728  & 0.261 \\
    BP95-ND     \cite{BP95}      & 0.726  & 0.268 \\
    P94-ND      \cite{P94}       & 0.726  & 0.270 \\
    RVCD96-ND   \cite{RCVD96}    & 0.725  & 0.278 \\
    BCDSTT      \cite{Basu}      & 0.721  & 0.279  \\
    TCL         \cite{TCL}       & 0.725  & 0.271 \\
\end{tabular}
\label{tab2}
\end{table}

%\begin{table}[htb]
%\centerline{\bf Table}
%\caption[tt]{
%Properties of the convective zone. The range of helioseismological estimates,
%the predictions of our SSM and of the model
%with 10\% opacity reduction.
%}
%\centering
%~\\
%\begin{tabular}{lccc}
%\hline
%\hline
%         &   Helios.     &  SSM    & $x=0.9$\\
%\hline
%$R_b/R_\odot$    & 0.710 -- 0.716  &  0.7158 & 0.7164 \\
%$c_b$[Mm/s]& 0.221 -- 0.225&  0.2223 & 0.2219  \\
%$Y_{photo}$&0.233 -- 0.268   &  0.238  & 0.223  \\
%\hline
%\hline
%\end{tabular}
%\label{tab1}
%\end{table} 

%\vskip1.5cm

%\centerline{\bf Figure captions}

\begin{thebibliography}{999}
\bibitem{Ga}
For the Gallium signal we use the weighted average of Gallex \cite{Gallex}
and Sage \cite{Sage} results.
\end{thebibliography}
\end{document}